\documentclass[12pt]{article}
\usepackage{amsthm,amsfonts,amsmath, amscd}

\swapnumbers
                              
\theoremstyle{plain}

\numberwithin{equation}{section}
\newtheorem{theorem}{THEOREM}[section]

\theoremstyle{definition}

\newtheorem{problem}[theorem]{PROBLEM}

\newtheorem{conjecture}[theorem]{CONJECTURE}

\theoremstyle{remark}

\numberwithin{equation}{section}

\newcommand{\bR}{{\mathbb R}}
\newcommand{\bH}{{\textbf{H}}}

\newcommand{\cH}{{\mathcal H}}

\newcommand{\cE}{{\mathcal E}}

\newcommand{\cX}{{\mathcal X}}

\newcommand{\ket}[1]{\left\vert #1\right\rangle}
\newcommand{\bra}[1]{\left\langle #1\right\vert}

\newcommand{\Tr}{\mbox{Tr}}

\newcommand{\be}{\begin{equation}}
\newcommand{\ee}{\end{equation}}
\newcommand{\bea}{\begin{eqnarray}}
\newcommand{\eea}{\end{eqnarray}}
\newcommand{\beann}{\begin{eqnarray*}}
\newcommand{\eeann}{\end{eqnarray*}}

\usepackage{color}

\begin{document}   
\title{Upper bounds on mixing rates}
\author{
\vspace{5pt}  Elliott H. Lieb$^{1}$  and Anna Vershynina$^{2}$ \\
\vspace{5pt}\small{$^{1}$ Departments of Mathematics and Physics, Jadwin
Hall, Princeton University, Princeton, NJ 08544 }\\
\vspace{5pt}\small{$^{2}$ Department of Physics, Jadwin
Hall, Princeton University, Princeton, NJ 08544}
}



\date{\today}

\footnotetext
[12]{Work partially
supported by U.S. National Science Foundation
grant PHY 0965859  and a grant from the Simons Foundation (\# 230207 to
Elliott Lieb).  \\
\copyright\, 2013 by the authors. This paper may be
reproduced, in its
entirety, for non-commercial purposes.}

 \maketitle

\begin{abstract}
We prove upper bounds on the rate, called "mixing rate", at which the von 
Neumann entropy of the expected density operator of a given ensemble of states changes under non-local unitary evolution.
For an ensemble consisting of two states, with probabilities of $p$ 
and $1-p$, we prove that the mixing rate is bounded above by $4\sqrt{p(1-p)}$ for any Hamiltonian of norm $1$. For a general ensemble
 of states
with probabilities distributed according to a random variable $X$ and individually evolving according to any set of bounded Hamiltonians,
 we conjecture that the mixing rate is bounded above by a Shannon entropy
of a random variable $X$. For this general case we prove an upper bound that is independent of the dimension of the Hilbert space on which 
states in the ensemble act. 
\end{abstract}

\vspace*{10pt}



\section{Introduction}

The problem addressed in this paper is, given an ensemble of states, $\cE$, to find an upper bound on the rate, $\Lambda(\cE)$,
 at which the von 
Neumann entropy of the expected density operator of this ensemble changes under non-local unitary evolution. 
The conjecture is known as 'Small Incremental Mixing' for an ensemble consisting of two states, $\cE_2$, and it states that the 
mixing rate is bounded above
by a binary entropy $S(p)=-p\ln p-(1-p)\ln(1-p)$, where $p$ and $1-p$ are the probabilities of the two states in the ensemble. 
The problem, to our knowledge, was first introduced by Bravyi in \cite{B07}.

We prove, Theorem \ref{Bin_Thm}, that for any ensemble consisting of two states, the mixing rate is bounded above by the following constant, which is independent of the dimension of the Hilbert space these states act on (including infinite dimension):
$$\Lambda(\cE_2)\leq 4\sqrt{p(1-p)}.$$ This bound has a shape similar to that of the binary entropy, which appears in the conjecture, up to a factor of 2.
But, unfortunately, our $\sqrt{p}$ behavior near $p=0$ is significantly worse than $p\ln p$. 

Bravyi proved \cite{B07} the Small Incremental Mixing conjecture for a special case in which the 
expected density operator has at most two distinct eigenvalues,
of arbitrary multiplicity. He could bound the mixing rate by $6$ times the binary entropy $S(p)$. 
For a general case he gave the dimension-independent and $p$-independent bound of $2$, which was the best dimension-independent bound
 until now. 
See Section 2 for the discussion of this conjecture.

In our paper, see Section 3, we generalize the problem to the ensemble of any number of states, $\cE$, not only two, and conjecture that the upper bound 
should be a 
Shannon entropy of the random variable $X$, according to which the probabilities are distributed in the ensemble, 
$$\Lambda(\cE)\leq S(X).$$
We prove that the 
mixing rate has an upper bound
independent of the dimension of the Hilbert space the states act on.
See Theorem \ref{Main_Thm} for the formulation of the result and Section 4 for the proof of the upper bound.

Bravyi introduced the Small Incremental Mixing problem as a generalization of the 'Small Incremental Entanglement'
conjecture, \cite{B07}. According to Bravyi, the latter conjecture was first proposed by Kitaev 
in a private communications to him. It bounds the rate of change of an entanglement between two parties when the system evolves
 under a non-local unitary evolution. The conjecture states that the upper bound is $c\ln d$, where $d$ is a dimension of a system of either
party and $c$ is a constant independent of either dimension. See Section 2 for the discussion of this conjecture.

The question of bounding a mixing rate by a binary entropy for an ensemble of two states is still open. In the special case discussed 
by Bravyi \cite{B07}, one would hope to improve the constant $6$ in front of the binary entropy. A conjecture of bounding a mixing rate
by a Shannon entropy for a general ensemble is open as well.

The paper is organized as follows. In Section 2 we discuss original Small Incremental Mixing problem (for the ensemble consisting of 
two states), Small Incremental Entangling problem, the relation between the two and the progress on both problems. 
In Section 3 we generalize Small Incremental Mixing to a 
general ensemble consisting of any number of states, pose a new conjecture and provide our main result on the upper bound of the mixing
rate, Theorem \ref{Main_Thm}. Section 4 contains the proof of Theorem \ref{Main_Thm}.

\section{Preliminaries}

Let $\cH$ denote a $D$-dimensional Hilbert space (which could be infinite dimensional). Let $\cE_2=\{(p,\rho_1), (1-p, \rho_2)\}$ be a probabilistic ensemble of two states acting
on $\cH$. 
The expected density operator of this ensemble is a convex combination $\rho=p\rho_1+(1-p)\rho_2$. For any Hamiltonian $H$ (self-adjoint
operator on $\cH$) we can define
a time dependent state $$\rho(t)=p\rho_1+(1-p)e^{-iHt}\rho_2e^{iHt}.$$
That is $H$ acts locally on $\rho_2$, but not on $\rho_1$.

The von Neumann entropy of this state is $$S(\rho(t))=-\Tr\Bigl(\rho(t)\ln\rho(t)\Bigr).$$

From the basic properties of von Neumann entropy, the following holds.

\textbf{Small Total Mixing. (Binary case)}\textit{ For any fixed ensemble $\cE_2$, the entropy of a state $\rho(t)$ at any time $t$ satisfies
$$ \overline{S}(\cE_2)\leq S(\rho(t))\leq \overline{S}(\cE_2)+S(p),$$
where $\overline{S}(\cE_2)=pS(\rho_1)+(1-p)S(\rho_2)$ is the average entropy of the ensemble and 
$S(p)=-p\ln p-(1-p)\ln(1-p)$ is a binary entropy.}

The inequality is proved in Chapter 3 for a general ensemble of any number of states, see (\ref{STM}).

The  analogue of the small total mixing for infinitely small times is formulated in terms of 
a mixing rate.

A \textit{\textbf{mixing rate}} is defined as
\begin{equation*}\label{MR_bin}
 \Lambda(\cE_2, H)=\frac{dS(\rho(t))}{dt}\bigg|_{t=0}.
\end{equation*}

\begin{conjecture}\label{Con_B}\textbf{(Bravyi \cite{B07}) Small Incremental Mixing}.\\
\textit{For any ensemble $\cE_2=\{(p,\rho_1), (1-p, \rho_2)\}$, the maximum mixing rate is bounded above by a binary entropy.}
\begin{align*}
 \Lambda(\cE_2):&=\max\{|\Lambda(\cE, H)| :\|H\|=1\}\\
&\leq S(p)=-p\ln p-(1-p)\ln(1-p).\nonumber
\end{align*}
\end{conjecture}
Some useful formulas for the mixing rate are
\begin{align}\label{L_bin}
 \Lambda(\cE_2, H)&=-ip\Tr([\rho_1,\ln\rho]H)\\
&=i(1-p)\Tr([\rho_2,\ln\rho]H)\nonumber,
\end{align}
and
\begin{align*}
 \Lambda(\cE_2)&=p\Tr|[\rho_1,\ln\rho]|\\
&=p\|[\rho_1,\ln\rho]\|_1,
\end{align*}
here the maximum is achieved for $H=1-2R$, with $R$ being a projector on the negative eigenspace of $i[\rho_1,\ln\rho]$. 
The norm $\|\cdot\|_1$ is 
a trace-norm.

Bravyi \cite{B07} proved that $\Lambda(\cE_2)\leq 6S(p)$, where $\rho$ has at most two distinct eigenvalues of arbitrary 
multiplicity.

Our result for an ensemble of two states is the following theorem.

\begin{theorem}\label{Bin_Thm}\textbf{(Binary case)}
 For any binary ensemble $\cE_2=\{(p,\rho_1), (1-p, \rho_2)\}$, the maximum mixing rate is bounded above
\begin{equation*}
 \Lambda(\cE_2)\leq 4\sqrt{p(1-p)}.
\end{equation*}
\end{theorem}
The proof of this theorem is given in Chapter 3 for a more general case, when the ensemble consists of any number of states, see 
Theorem \ref{Main_Thm}.

Although we do not pursue this direction, we mention the following question posed by Audenaert and Kittaneh \cite{AK12}.

\begin{problem}\label{Con_AK}\textbf{(Audenaert, Kittaneh \cite{AK12})}
\textit{Let $A$ and $B$ be arbitrary positive semi-definite $D\times D$ matrices with $alpha=\Tr \, A$ and $\beta=\Tr\,  B$. For what functions 
$f:\bR\rightarrow \bR$
does there exist a constant $\gamma$, independent of $d$, $A$  and $B$, such that}
$$\|[B, f(A+B)]\|_1\leq \gamma(F(\alpha+\beta)-F(\alpha)-F(\beta)),$$
\textit{where $F(x)=\int_0^xf(y)dy$?}
\end{problem}
Note that if $A=p\rho_1$, $B=(1-p)\rho_2$ and $f(x)=\ln x$, Problem \ref{Con_AK} becomes Conjecture \ref{Con_B}.

As mentioned in the introduction, the Small Incremental Mixing problem is a generalization of the Small Incremental Entangling conjecture.
To formulate the later conjecture, we suppose that two parties, say Alice and Bob, have control over systems $A$ and $B$. Both
systems evolve according to a non-local Hamiltonian $H_{AB}$. In time entanglement between $A$ and $B$ can be generated.
In ancilla-assisted entangling both parties have access to additional subsystems, called local ancillas, i.e. Alice is in control of two systems $A$ and $a$ and
Bob is in control of $B$ and $b$. Alice and Bob start with a pure state $\rho(0)=\ket{\Psi}\bra{\Psi}$ on the system $aABb$.

A time dependent joint state of Alice and Bob is $$\rho(t)=U^*(t)\ket{\Psi}\bra{\Psi}U(t),$$ where $U(t)=I_a\otimes e^{iH_{AB}t}\otimes I_b$
is a unitary transformation. The joint state of Alice and Bob is pure at any time.

One of the ways to describe the entanglement between Alice and Bob is to calculate the entanglement entropy
$$E(\rho(t)):=S(\rho_{aA}(t))=-\Tr\rho_{aA}(t)\ln \rho_{aA}(t),$$
where $\rho_{aA}(t)=\Tr_{Bb}\rho(t)$ is a state that Alice has after time $t$. Since the joint state is pure, the entanglement entropy
also can be calculated from the state that Bob has $E(\rho(t))=S(\rho_{Bb}(t))$.

\textbf{Small Total Entangling.}\textit{ The total change of the entanglement $E(\rho(t))$ is at most $2\ln d$, where $d=\min\{\dim(A), \dim(B)\}$.} 
See 
\cite{BHLS03} for the proof.

A problem of bounding the infinitesimal change of the entanglement is formulated using the entangling rate.

The \textit{\textbf{entangling rate}} is defined by
$$ \Gamma(\Psi, H)=\frac{dE(\rho(t)}{dt}\bigg|_{t=0}.$$

After calculating the derivative, the entangling rate can be expressed as
\begin{align}\label{Ent_rate}
 \Gamma(\Psi, H)&=-i\Tr\Bigl(H_{AB}[\rho_{aAB}, \ln(\rho_{aA})\otimes I_B]\Bigr)\nonumber\\
&=-i\Tr\Bigl(H_{AB}[\rho_{aAB}, \ln(\rho_{aA}\otimes \frac{I_B}{\dim(B)})]\Bigr).
\end{align}
Similarly, $\Gamma(\Psi, H)=-i\Tr\Bigl(H_{AB}[\rho_{ABb}, I_A\otimes\ln(\rho_{Bb})]\Bigr) $

\begin{conjecture}\textbf{(Bravyi \cite{B07}) Small Incremental Entangling.}\\
 \textit{There is a universal constant $c$ such that for all dimensions of ancillas $a$, $b$ and for all states $\ket{\Psi}$, the 
following holds}
$$\Gamma(\Psi, H)\leq c \|H\|\ln d, $$
\textit{where $d=\min\{\dim(A), \dim(B)\}.$}
\end{conjecture}

This problem was studied by many authors. The case with no ancillas was proved by Bravyj \cite{B07} for the pure initial state and by Hutter and Wehner \cite{HW12},
\cite{H11} for either pure or mixed initial state.  For the case when $A$ and $B$ are quibits, Childs et al \cite{CLV06} give upper bounds for 
the entangling rate and show that they are independent of the ancillas $a$ and $b$. Wang and Sanders \cite{WS03} proved that
$\Gamma(H):=\max_{\Psi}\Gamma(\Psi, H)\leq\beta\|H\|$, where $\beta\approx 1.9123$, for an uncorrelated Hamiltonian $H=H_A\otimes H_B$, when $H_{A(B)}=H_{A(B)}^{-1}$.  Child et al \cite{CLV04} also proved an upper bound 
$\Gamma(H)=\beta\frac{1}{4}\Delta_A\Delta_B\leq \beta\|H\|$ for the ancilla-assisted case and for an
arbitrary uncorrelated bipartite Hamiltonian $H=H_A\otimes H_B$, where $\Delta_A$ ($\Delta_B$) is the difference between the largest and 
smallest
eigenvalues of $H_A$ ($H_B$). For an arbitrary bipartite Hamiltonian Bennet et al \cite{BHLS03} proved that 
$\Gamma(H)\leq c d^4\|H\|$, where $c$ does not depend on $a$ or $b$. 

Bravyi \cite{B07} proved that Small Incremental Mixing with a constant $c$ in front of the Shannon entropy implies Small Incremental 
Entangling with a constant $4c$, by choosing particular ensemble of states: 
$\cE_2=\{((1-\dim(B)^{-2}), \mu_{aAB}), \dim(B)^{-2}, \rho_{aAB})\}$. Here without loss of generality it was assumed that $B\leq A$ and
 $\mu_{aAB}$ is a state such that the expected density operator 
of the ensemble is of the form appearing in (\ref{Ent_rate})
\begin{equation}\label{Implementation}
\rho_{aA}\otimes\frac{I_B}{B}=\Bigl(1-\dim(B)^{-2}\Bigr)\mu_{aAB}+\dim(B)^{-2}\rho_{aAB}.
\end{equation}
In Lemma 1 \cite{B07} it was proved that such a state $\mu_{aAB}$ exists.
Applying (\ref{L_bin}) to this ensemble, we have that $\Lambda(\cE_2, H)=\dim(B)^{-2}\Gamma(\Psi, H)$, which shows that Small
Incremental Mixing implies Small Incremental Entangling.

Using Bravyi's proof, our bound of $4\sqrt{p(1-p)}$ for the Small Incremental Mixing problem leads to a bound
of $4d\|H\|$ in Small Incremental Entangling.

\section{Small Incremental Mixing for an ensemble consisting of any number of states}

We generalize the Small Incremental Mixing 
problem to an ensemble of any number of states in the following way.

Let $X$ be a random variable with probability density $p_X(x)$, i.e. the probability 
that the realization $x$ occurs is $p_X(x)$, where the realization $x$ belongs to a set $\mathcal{X}$. 
Consider a probabilistic ensemble of states $\mathcal{E}=\{p_X(x), \rho_x\}_{x\in\cX}$, i.e. $\rho_x$ is a density matrix on a Hilbert space $\cH$ of arbitrary dimension (including infinite dimension), which occurs 
with probability $p_X(x)$, where $\sum_x p_X(x)=1$. 

The expected density operator of the ensemble $\mathcal{E}$ is convex combination of density matrices $\rho=\sum_xp_X(x)\rho_x$. 
For any collection of Hamiltonians 
$\bH=\{H_x\}_{x\in\cX}$ define a time-dependent state $$\rho(t)=\sum_{x\in\cX} p_X(x)e^{-iH_xt}\rho_xe^{iH_xt}.$$
Note that one of the states could always be left invariant, i.e. one of the Hamiltonians could always be taken as an identity $I$, but 
to simplify the notation we shall write a time evolution for all states.

The von Neumann entropy $S(\rho(t))$ of this state satisfies the following property.

\textbf{Small Total Mixing (General case).}\textit{ For any fixed ensemble $\cE$, the entropy of a state $\rho(t)$ at any time $t$ satisfies
\begin{equation}\label{STM}
 \overline{S}(\cE)\leq S(\rho(t))\leq \overline{S}(\cE)+S(X),
\end{equation}
where $\overline{S}(\cE)=\sum_xp_X(x)S(\rho_x)$ is the average entropy of an ensemble $\cE$ and $S(X)=-\sum_xp_X(x)\ln p_X(x)$ is 
a Shannon entropy of a classical random variable $X$.}

The lower bound follows from concavity property of the von Neumann entropy and the invariance of the entropy under unitary 
transformation 
\begin{align*}
\overline{S}(\cE)=\sum_xp_X(x)S(\rho_x)&=\sum_xp_X(x)S(e^{-iH_xt}\rho_xe^{iH_xt})\\
&\leq S\Bigl(\sum_{x\in\cX} 
p_X(x)e^{-iH_xt}\rho_xe^{iH_xt}\Bigr)=S(\rho(t)).
\end{align*}

To see the upper bound, form a classical-quantum state $$\rho^{XA}(t)=\sum_xp_X(x)\ket{x}\bra{x}^X\otimes\rho_x(t),$$ that acts on a tensor product of a
classical space $X$ and a quantum system $A$ represented by a Hilbert space $\cH$. Here, $\rho_x(t)=e^{-iH_xt}\rho_xe^{iH_xt}.$
The entropy of the classical-quantum state is $$S(\rho^{XA}(t))=S(X)+\overline{S}(\cE).$$ For a classical-quantum state $\rho^{XA}$ the relative entropy
$S(X|A):=S(\rho^{XA})-S(\rho^A)\geq 0$ is always non-negative, therefore $S(\rho(t))\leq S(\rho^{XA}(t)).$ This proves the Small Total 
Mixing property.

The  analogue of the small total mixing for infinitely small times is formulated in terms of 
a mixing rate.

A \textit{\textbf{mixing rate}} is defined similarly to the binary case as
\begin{equation}\label{MR}
 \Lambda(\cE, \bH)=\frac{dS(\rho(t))}{dt}\bigg|_{t=0}.
\end{equation}

\begin{conjecture}\textbf{Small Incremental Mixing}.\\
\textit{For any ensemble $\cE=\{(p_X(x),\rho_x)\}_{x\in\cX}$, the maximum mixing rate is bounded above by a Shannon entropy.}
\begin{align}
 \Lambda(\cE):&=\max\{|\Lambda(\cE, \bH)| :-I\leq H_x\leq I, x\in\cX\}\label{def_Lmb}\\
&\leq S(X)=-\sum_xp_X(x)\ln p_X(x).\nonumber
\end{align}
\end{conjecture}

A mixing rate can be written explicitly by calculating a derivative of the entropy
$ \frac{dS}{dt}(\rho(t))=-\Tr\Bigl(\frac{d\rho(t)}{dt}\ln\rho(t)\Bigr),$ at $t=0$
\begin{align}
\Lambda(\cE,\bH)&=-i\sum_xp_X(x)\Tr([H_x,\rho_x]\ln\rho)\nonumber\\
&=-i\sum_xp_X(x)\Tr(H_x[\rho_x,\ln\rho]).\label{Lambda}
\end{align}

Note that in the definition of the mixing rate (\ref{def_Lmb}) the maximum is taken over all Hamiltonians $H_x$ such that $-I\leq H_x\leq I$.

For any Hermitian operator $A$ with $\Tr(A)=0$, $$\max\{\Tr(HA): -I\leq H\leq I\}=2\max\{\Tr(HA): 0\leq H\leq I\}.$$ 
This property can be
easily observed by expressing $H=2R-I$, where $0\leq R\leq I$. 

Therefore the maximum in (\ref{def_Lmb}) can be taken over the non-negative 
Hamiltonians bounded above by identity operator.
\begin{align}\label{L(p)}
&\Lambda(\cE)=2\max\{|\Lambda(\cE,\bH)|: 0\leq H_x\leq I, x\in\cX\}.
\end{align}
Note that, similarly to the binary case, the maximum is achieved for the Hamiltonians $H_x$ being a 
projector onto a positive eigenspace of $i[\rho_x,\ln\rho].$

Bravyi's proof of the Small Incremental Mixing problem for an ensemble of two states and for a state $\rho$ with binary spectrum can be 
easily generalized to the case of an ensemble containing any number of states with $\rho$ still restricted to having a binary spectrum. 
Also the constant $6$ in front of the Shannon entropy $S(X)$ remains.

In Section 3 we prove the following theorem, showing that the maximum mixing rate is bounded above
by a constant independent of the dimension $D$ of the Hilbert space $\cH$ that states act on.
\begin{theorem}\label{Main_Thm}\textbf{(General case)}
 For a fixed ensemble $\cE=\{p_X(x),\rho_x\}_{x\in\cX}$ the maximum mixing rate (\ref{def_Lmb}) is bounded above
\begin{equation*}
 \Lambda(\cE)\leq 4\sum_{x\neq x_0}\sum_{y\neq x}\sqrt{p_X(x)p_X(y)},
\end{equation*}
where $x_0\in\cX$ such that $p_X(x_0)$ is the largest among $x\in\cX$.
\end{theorem}
For a binary ensemble $\cE=\{(p, \rho_1), ((1-p), \rho_2)\}$ Theorem \ref{Main_Thm} gives the upper bound of $4\sqrt{p(1-p)}$, as 
claimed in Theorem \ref{Bin_Thm}.

\section{Upper bound on the mixing rate}
In this section we prove Theorem \ref{Main_Thm}.

In eq. (\ref{Lambda}), express the logarithm $\log\rho$ by the formula $$\ln x=\int_0^\infty\Bigl(\frac{1}{1+t}-\frac{1}{x+t}\Bigr)dt.$$
We may assume that $0\leq H_x\leq I$ for every $x\in\mathcal{X}$, as noted before (\ref{L(p)}), to calculate the mixing rate.
As noted in the construction of $\rho(t)$ one may take one of the Hamiltonians $H_x$ to be equal to the identity.
Take $H_{x_0}=I$, where $x_0$ is such that $p_X(x_0)$ is the largest among $x\in\cX$.
Then, for reasons explained below,
\begin{align*}
 \Lambda(\cE,\bH)&=-i\sum_{x\neq x_0}p_X(x)\Tr(H_x [\rho_x,\ln\rho])\\
&=i\sum_{x\neq x_0}p_X(x)\int_0^\infty\Tr\Bigl(H_x [\rho_x,\frac{1}{\rho+t}] \Bigr)dt\\
&=-i\sum_{x\neq x_0}p_X(x)\int_0^\infty\Tr\Bigl(H_x\frac{1}{\rho+t} [\rho_x,\rho+t]\frac{1}{\rho+t} \Bigr)dt\\
&=-i\sum_{x\neq x_0}\sum_{y\neq x}p_X(x)p_X(y)\int_0^\infty\Tr\Bigl(H_x\frac{1}{\rho+t} [\rho_x,\rho_y]\frac{1}{\rho+t} \Bigr)dt\\
&=-i\sum_{x\neq x_0}\sum_{y\neq x}p_X(x)p_X(y)\int_0^\infty\Tr \Bigl( [\rho_x,\rho_y]\frac{1}{\rho+t}H_x\frac{1}{\rho+t}\Bigr)dt.
\end{align*}
Here in the third equality we used that $[A, {B}^{-1}]=-B^{-1}[A, B]B^{-1}.$ In the fourth equality we wrote $\rho$ as 
a convex combination of states $\rho_y$ and eliminated the commuting terms. In the last equality the cyclicity of the trace is used.
 
For any $0\leq H\leq I$, we have $0\leq (\rho+t)^{-1}H(\rho+t)^{-1}\leq (\rho+t)^{-2}$.

Therefore, continuing our calculations, with $K_x:=\int_0^\infty(\rho+t)^{-1}H_x(\rho+t)^{-1}dt\leq \rho^{-1}$, $x\in\cX$
\begin{align*}
| \Lambda(\cE,\bH)|&=\Bigl|\sum_{x\neq x_0}\sum_{y\neq x}p_X(x)p_X(y)\Tr \Bigl( [\rho_x,\rho_y]\int_0^\infty\frac{1}{\rho+t}H_x\frac{1}{\rho+t}dt\Bigr)\Bigr|\\
&\leq \sum_{x\neq x_0}\sum_{y\neq x}p_X(x)p_X(y)\Bigl(|\Tr\rho_x\rho_y K_x|+|\Tr\rho_y\rho_x K_x|\Bigr)\\
&\leq \sum_{x\neq x_0}\sum_{y\neq x}p_X(x)p_X(y)\Bigl(\Tr|\sqrt{K_x}\rho_x\rho_y\sqrt{K_x}|+\Tr|\sqrt{K_x}\rho_y\rho_x\sqrt{K_x}|\Bigr)\\
&\leq 2\sum_{x\neq x_0}\sum_{y\neq x}p_X(x)p_X(y)\sqrt{\Tr(\rho_x^2K_x)\Tr(\rho_y^2K_x)}\\
&\leq 2\sum_{x\neq x_0}\sum_{y\neq x}\sqrt{p_X(x)p_X(y)}\sqrt{\Tr\Bigl(\rho_x(p_X(x)\rho_x)\rho^{-1}\Bigr)\Tr\Bigl(\rho_y(p_X(y)\rho_y)\rho^{-1}\Bigr)}\\
&\leq 2\sum_{x\neq x_0}\sum_{y\neq x}\sqrt{p_X(x)p_X(y)}.
\end{align*}
In the first inequality we put the absolute value inside the sums, wrote the commutator $[\rho_x, \rho_y]=\rho_x\rho_y-\rho_y\rho_x$ and
 used the triangular property of the absolute value. The second inequality follows from the cyclicity of the trace and by moving the absolute value inside the trace.
The third inequality follows from Cauchy-Schwartz inequality for traces: $$\Tr|AB|\leq \sqrt{\Tr(A^*A)\Tr(B^*B)}.$$ 
The fourth inequality follows from the upper bound on $K_x\leq \rho^{-1}$.
The fifth inequality is obtained from the definition of $\rho$ as a convex combination of non-negative density operators $\rho_x$, therefore
$\rho\geq p_X(x)\rho_x$ for any $x\in\cX$, and $\Tr\rho_x=1$, $x\in\cX$.

From (\ref{L(p)}) we obtain an upper bound for the mixing rate
\begin{align*}
 \Lambda(\cE)&=2\max_{0\leq H_x\leq I}\{|\Lambda(\cE, \bH)|\}\leq 4\sum_{x\neq x_0}\sum_{y\neq x}\sqrt{p_X(x)p_X(y)}.
\end{align*}
\qed

\textbf{Acknowledgments.} We are grateful to Frank Verstraete for making us aware of the mixing rate problem and for his encouragement.

\end{document}